# Diagnostics of atmospheric pressure capillary DBD oxygen plasma jet

N. C. Roy, M. R. Talukder[a)], and B. K. Pramanik[b)]

*Plasma Science and Technology Laboratory, Department of Applied Physics and Electronic Engineering,*

*[b)]Department of Computer Science and Engineering,University of Rajshahi, Rajshahi 6205, Bangladesh.*

*[a)]Author to whom correspondence should be addressed, Electronic mail: mrtalukder@ru.ac.bd.*

Atmospheric pressure capillary dielectric barrier oxygen discharge plasma jet is developed to generate non-thermal plasma using unipolar positive pulse power supply. Both optical and electrical techniques are used to investigate the characteristics of the produced plasma as function of applied voltage and gas flow rate. Analytical results obtained from the optical emission spectroscopic data reveal the gas temperature, rotational temperature, excitation temperature and electron density. Gas temperature and rotational temperature are found to decrease with increasing oxygen flow rate but increase linearly with applied voltage. It is exposed that the electron density is boosting up with enhanced applied voltage and oxygen flow rate, while the electron excitation temperature is reducing with rising oxygen flow rate. Electrical characterization demonstrates that the discharge frequency is falling with flow rate but increasing with voltage. The produced plasma is applied preliminarily to study the inactivation yield of *Fusarium oxysporum* infected potato samples.

## I. INTRODUCTION

In the recent years, atmospheric pressure dielectric barrier discharge plasma jets (APDJ) have drawn great interest to the researchers due to its application in biology[1], chemical synthesis, polymer and material surface modifications[2-6]. The APDJs produce a large number of ions, reactive species and free radicals because the energetic electrons enhance the gas chemistry. APDJs are able to generate nonthermal plasma, with low gas temperature and high electron temperature, in open space rather than in confined discharge chamber without any arrangement for controlling pressure inside the chamber. Another advantage is that the size and position of the sample to be treated is not a factor and the plasma jet can be applied on the sample directly. Several parameters can control the properties[7,8] of APDJ such as jet geometry, type of dielectric, excitation frequency, type of power source and gas composition including gas flow rate. Due to low gas temperature, APDJs do not cause thermal damage to heat-sensitive biological[9] systems such as living tissues and cells. These application processes can be enhanced and controlled by better understanding the properties of the produced plasma. Because, the gas flow dynamics are different[7] in each type of jet due to jet structure with nozzle geometry, Reynolds number and initial velocity. The objective of this work is to develop and investigate the properties of the produced plasma jet for biological applications.

Gas temperature ($T_g$), electron density ($n_e$) and electron temperature ($T_e$) are the three fundamental properties of plasmas and have their significant effects on the operation and maintenance of gas discharge. In most biological applications[1] of APDJs, $T_g$ has a significant influence[10] on the sample to be investigated, so that $T_g$ needs to diagnose more precisely in order to attain at room temperature and maintain highly stable for efficient chemical reactions to meet the pertaining requirements. Electrical and optical characterizations are carried out for understanding the properties of the APDJ.

Experimental setup for the production and measurements of APDJ plasma is presented in section II. Analytical results and discussion concerned with the measurements are discussed in section III. Finally conclusion is drawn in section IV.

## II. EXPERIMENTAL SETUP

The schematic diagram of the experimental setup of APDJ is shown in Fig.1. It consists of a high voltage power source, plasma jet, and oxygen ($O_2$) gas cylinder incorporating with a gas flow controller. The power supply used for the production of plasma is designed and constructed in our laboratory. The open circuit voltage, short circuit current and frequency are approximately $10\ kV$, $160mA$ and $60\ kHz$, respectively. In the design it is ensured that the power supply can operate both in unipolar positive and bipolar modes as necessary. The jet is made using three pieces of pyrex glass tubes. Outer diameters of the outer and inner glass tubes are $15mm$ and $7mm$, respectively. Length, outer and inner diameters of the capillary tube are $15mm$, $1.4mm$, and $0.9mm$, respectively. The inner electrode, made of a $0.50mm$ diameter copper wire, is placed inside the capillary tube. A cylindrical outer electrode, made of copper tube of $1.4mm$ inner diameter, is placed outside the capillary tube. Spacing between the inner electrode and the inner surface of the capillary tube is maintained at $0.2\ mm$. The inner and outer electrodes are connected with the high voltage power supply as shown in Fig. 1. In order to prevent arc from the surface of the outer electrode, the space between the outer and inner glass tube is filled with transformer oil. The electrodes are arranged such that the cross-electric field can prevail within the electrode gap. Unipolar positive pulse voltage is applied across the electrodes. Our experimental setup is different from that of Thiyagarajan *et al.*[11] in that they used linear-field type electrode configuration with bipolar sinusoidal type voltage source having different source parameters. In our setup, commercial grade $O_2$ gas (99.65%) is fed through the upper side of the inner glass tube from the $O_2$ gas cylinder. The flow rate of $O_2$ gas is controlled by a gas flow controller ($KIT\ 115P$). The voltage across the electrodes is measured with an oscilloscope ($GDS$ 1022) in combination with a high voltage probe ($HVP - 08$). The current flowing through the plasma is measured by a current probe ($CP - 07C$). Emitted spectrum from the produced plasma is fed to the spectrophotometer (USB2000 + XR1) through a 200 $cm$ long optical fiber cable. The spectrophotometer is associated with a computer for spectral data acquisition. At different applied voltages and $O_2$ gas flow rates, the corresponding discharge voltages, currents and emitted spectra are recorded in order to study the properties[12-14] of the APDJ discharges.

Typical discharge voltage and current waveforms of the APDJ measured at 2.5 $kV$ with $O_2$ flow rate of 3 $lpm$ are displayed in Fig. 2. The recorded emission spectrum along with the identified major peaks of the species is depicted in Fig. 3. The analytical results are discussed in the following sections.

## III. RESULTS AND DISCUSSION

### A. Electrical characterization

The voltage is applied to the inner electrode keeping the outer electrode floating. $O_2$ gas is injected into the capillary jet. With increasing the applied voltage to some extent, breakdown of $O_2$ gas is occurred. The production and maintenance of the capillary $O_2$ DBD plasma jet are found to be sensitive to the applied voltage and gas flow rate. The electronegative $O_2$ plays a significant role in plasma discharge in air[15]. Generally, breakdown of $O_2$ occurs from several micro-discharges that are randomly spread in time and space at atmospheric pressure DBD plasma jet[3,6,11,16]. Approximately uniform DBD plasma is produced with $O_2$ gas. At the starting phase of the discharge, the discharge current delays approximately 0.80 $\mu s$ the discharge voltage in each discharge cycle. It is seen that the discharge current starts to flow when the applied voltage pulse arrives to a certain voltage level. Energetic electrons, because of gaining kinetic energy from the applied electric field, excite oxygen species to higher energy levels through inelastic collisions during the drifting process of the electrons. When the applied electric field is greater than the critical value of breakdown, electrons move toward the power electrode and hence the electron density increases through electron avalanche process[17]. However, the discharge current flows as soon as the conduction channel is formed in the discharge gap and hence the discharge voltage approaches near zero across the electrode gap. This process repeats in each cycle of the discharge voltage pulse.

With increasing $O_2$ flow rate, the discharge volume and the luminosity of the discharges become apparently increase as shown in Fig. 4. The photographs presented are taken by $Canon\ EOS$ 1100$D$ digital camera and the exposure time for all images is fixed at 1/100 s. It is seen from the figure that for a constant voltage of 2.5 $kV$, the plasma discharge volume increases with growing gas flow rate. It is also clear from the images that the discharges appear intense within the electrode space and much weaker outside the electrode region. This occurs in such a cross-field capillary DBD jet plasma discharge due to the monotonic decay of electric field results from spatially confined gas ionization in the electrode region. The applied voltage with rapid oscillation provides radially directed momentum to the electrons, and hence it is difficult for electrons to be transported axially toward the downstream[18] of the capillary tube.

The operating frequency plays a significant role for the formation and maintenance of the discharge at atmospheric pressure DBD plasma jet. Fig. 5 shows the effect of gas flow rate and voltage on the discharge frequency. The discharge frequency spectrum is determined by taking Fourier transformation of the discharge current, as shown in the inset. The frequency spectrum indicates that the atmospheric pressure DBD plasma consists of many micro-discharges as mentioned earlier. However, in the plot, it is taken into account only that frequency which contains highest amplitude. The discharge frequency of the produced plasma changes with the change in applied voltage and gas flow rate. The spatial charges are accumulated to the power electrode. The drifting velocity of the spatial charge is related to the electric field. With increasing electric field, the drifting velocity of the spatial charges is also increased and hence the discharge frequency increases with the increase of applied voltage[17]. This phenomenon is counteracting the increased $O_2$ gas flow rate with discharge frequency. This can be attributed with the Reynolds number $R_e = n_g v_g l_c / \mu_d$, where $n_g$, $v_g$, $l_c$, and $\mu_d$ are the gas density, gas velocity, characteristic length and dynamic viscosity, respectively. The Reynolds number increases with increasing $O_2$ gas flow rate[7]. It is seen from the above relation that the density of $O_2$ gas at the exit of the jet nozzle increases with increasing the Reynolds number and hence the electric field threshold for $O_2$ gas breakdown increases. As a result, the discharge frequency decreases with increasing $O_2$ gas flow rate.

### B. Optical characterization

For plasma diagnostics, Langmuir probe at atmospheric pressure[19,20] provides better accuracy. But, due to small plasma volume and the structure of the jet in our experiment, the probe technique is not suitable for the diagnostics of the produced plasma. Hence, the optical emission spectroscopic (OES) technique is applied to analyze the plasma. The basis of OES data analyses for the determination of $T_g$ and $n_e$ are the line broadenings due to Doppler and Stark effects. While, molecular rotational temperature ($T_{rot}$) and electronic excitation temperature ($T_{exc}$) are determined using Boltzmann plot method. In order to have an in-depth understanding of APDJ, characteristics of the discharges are studied as function of voltage applied and gas flow rate to find out suitable experimental conditions for the expected biological application. Especially the gas temperature and the rotational temperature are estimated and compared, respectively, by using the broadenings of fine structure of $H_\beta$ line and the emission band of $OH$ radical.

## 1. Species identification

The species produced in the plasma are identified using the NIST database[21] and the plasma parameters are estimated analyzing the spectra. In Fig. 3, a typical OES spectrum, measured at $2.5\,kV$ with $O_2$ flow rate of $3\,lpm$ in air, is presented. The dominant peaks observed in the spectra are $NO\ (B^2\Pi - X^2\Pi)$ at $248.78\,nm$, $OH\ (X^2\Pi(v'=0) \rightarrow A^2\Sigma^+(v''=0))$ at $308.17$, $O^+\big((\,^3P)5f - (\,^3P)3d\big)$ at $320.06\,nm$, $O^+\big((\,^3P)3d - (\,^3P)3p\big)$ at $348.59\,nm$, $O^+\big((\,^3P)3d - (\,^3P)3p\big)$ at $374.42\,nm$, $O((\,^4S°)3s - (\,^4P°)3p)$ at $395.19\,nm$, $O^+((\,^5S°)3d - (\,^5S°)3p)$ at $414.37\,nm$, $O_2^+(A^2\Pi_u(v''=11) \rightarrow X^2\Pi_g(v'=2))$ at $439.80\,nm$, $H_\beta(4d - 2p)$ at $486.12\,nm$, $O^+((\,^4S°)5d - (\,^4S°)3p)$ at $533.07\,nm$, $O^+((\,^1S)3p - (\,^3P)3s)$ at $561.10\,nm$, $O^{3+}((\,^3P)3d - (\,^3P)3p)$ at $588.24\,nm$, $O^{3+}((\,^3P)4d - (\,^3P°)3s)$ at $614.23\,nm$, $O^{2+}(\,^2P°)4d - (\,^2P°)4p$ at $707.91\,nm$, $O((\,^5P)3p - (\,^5S)3s)$ at $776.06\,nm$, and $Ar((\,^3P)4p - (\,^1D)3d)$ at $794.08\,nm$. The emission from $OH$ radical can be originated from the fragmentation of water molecules present in the air. The emission band of $OH$ radical is shown in the inset of Fig. 3. Four peaks $R1(1)$, $P1(1)$, $P1(2)$ and $P1(3)$ of $OH$ radical are found at $307.20\,nm$, $308.15\,nm$, $308.63\,nm$, and $309.96\,nm$, respectively. It is to be noted that $O_2$ is used as working gas for plasma production in open air. The emission from $NO$ is also produced due to the interactions of the plasma species concerned with the air molecules. On the other hand, the presence of $Ar$ emission line in the discharge may be due to the existence of residual $Ar$ impurities in the $O_2$ gas. It is found from Fig. 3 that the emission lines of $O$ atoms at $395.19\,nm$ and $776.06\,nm$ contain high intensity in the spectra. This indicates that the density of $O$ atoms is much higher than any other species observed. Because the $O$ atoms are produced dominantly through $e + O_2 \rightarrow 2O + e$ and $e + O_2^+ \rightarrow 2O$ collisions due to their highest production rate coefficients[22,23] of $4.2 \times 10^{-9}\exp(-5.6/T_e)\ cm^3.s^{-1}$ and $5.2 \times 10^{-9}.T_e^{-1}\ cm^3.s^{-1}$, respectively, with any other species production rate coefficients those involved in the collision processes. For the determination of gas temperature, rotational temperature, excitation temperature and electron density using the concerned peaks and $OH$ bands are discussed below.

## 2. Gas temperature and rotational temperature

It is well known that the broadenings of emission line profiles are the consequences from the contribution of different interaction mechanisms of the particles concerned. The Doppler broadening ($\Delta\lambda_{Doppler}$) and the instrumental broadening ($\Delta\lambda_{inst}$) produce a profile that appears to as the Gaussian profile ($G(\lambda)$). On the other hand, the natural ($\Delta\lambda_{nat}$), resonance ($\Delta\lambda_{res}$), Van der Waals ($\Delta\lambda_{VdW}$), and Stark ($\Delta\lambda_{Stark}$) broadenings can be approximated by Lorentzian profiles[24] ($L(\lambda)$). Thus, each broadened line contains Doppler and Stark broadenings so that the Voigt function ($V(\lambda)$) is the convolution of the Gaussian and Lorentzian functions[8]. However, $T_g$ is determined from the full width at half maximum (FWHM) ($\Delta\lambda_G$) of the Gaussian profile by the following equation[25]

$$\Delta\lambda_{Doppler} = 7.16 \times 10^{-7} \cdot \lambda_o \sqrt{\frac{T_g}{M}} \ (nm), \qquad (1)$$

where $\lambda_o$ is central wavelength $(nm)$, $M$ is the atomic mass $(g.mol^{-1})$ of the emitter and $T_g$ is the gas temperature $(K)$. In our analyses, it is found that the contribution of $\Delta\lambda_{inst}$ is on average $12\%$ to $\Delta\lambda_G$ and hence it is neglected.

Sometimes, the rotational temperature is also considered as the gas temperature[26] of the atmospheric pressure plasma. But, Ionascut-Nedelcescu *et al*. pointed out that the gas temperatures determined using $H_\beta$ line and $OH$ band are different. Since, it is important to infer the appropriate gas temperature for biological application of the APDJ, as mentioned earlier. The rotational temperature can be estimated under the condition of low rotational levels of $OH(A^2\Pi(v^{'} = 0) - X^2\Sigma^+(v^{''} = 0))$ with lower rotational quantum number $J^{''}$. It is assumed that there is minimum self-absorption ($< 1\%$), negligible interferences with other species present in the plasma and the rotational levels of the upper vibrational state $v^{'}$ follow Boltzmann distribution with the rotational and translational degrees of freedom in the plasma that is valid in atmospheric pressure discharges. Under these conditions, the emission intensity of the rotational transition ($J^{'} \rightarrow J^{''}$) in a given band ($v^{'} = 0 \rightarrow v^{''} = 0$) can be expressed[26] as

$$I_{J^{'}J^{''}} = \frac{K}{{\lambda_{J^{'}J^{''}}}^4} S_{J^{'}J^{''}} exp\left(-\frac{E_{J^{'}}}{k_B T_r}\right), \qquad (2)$$

where $I_{J^{'}J^{''}}$, $\lambda_{J^{'}J^{''}}$, $S_{J^{'}J^{''}}$ and $E_{J^{'}}$ are the intensity, transition wavelength ($J^{'} \rightarrow J^{''}$), Hönl-London factor, the energy of upper level with rotational quantum number $J^{'}$, respectively and $K$ is a constant includes all the terms that are independent of rotational quantum number. The line parameters used for the determination of $T_{rot}$ are collected from Chang *et al*. The Boltzmann plot, shown in Fig. 7, for $ln(I_{J^{'}J^{''}}{\lambda_{J^{'}J^{''}}}^4/S_{J^{'}J^{''}})$ as a function of $E_{J^{'}}$ is drawn inserting the required line parameters in Eq. (2) along with the measured intensity $I_{J'J''}$ of the spectrum. The rotational temperature can be determined from the slope of linear fit.

In the present study, emission of $H_\beta(4d - 2p)$ line at $486.12\ nm$ and $OH$ band from $306 - 312nm$ are used for the determination of $T_g$ and $T_{rot}$, respectively. The upper and lower limits of temperatures are estimated from the data provided by the fitting software and mentioned within the bracket noted in Fig. 7. The estimated errors are found in the range from $15\% - 25\%$. In atmospheric pressure low temperature plasmas containing even a small amount ($\sim 1\%$) of water molecule produces $OH(A - X)$ transition. But the emission intensity of the $OH$ radical is very low due to strong dependence[27, 28] of intensity on the $T_{rot}$. However, the $T_{rot}$ is estimated from the relative intensities corresponding to the $R$ and $P$ branches of the $OH\ A - X\ (0,0)$ band with four distinct peaks of the two branches as shown in the inset of Fig. 3.

Figure 8 shows the effect of gas flow rate on $T_g$ and $T_{rot}$ at different discharge voltages. It is observed that $T_g$ and $T_{rot}$ are decreasing with increasing gas flow rate. Gas flow plays a significant role in carrying Joule heat[15] out from the discharge region. The reason of decreasing $T_g$ and $T_{rot}$ arises due to the fact that the transit time of the oxygen species becomes shorter within the electrode gap where strong electric field exists. The electrons gain energy from the electric field in this gap and releasing energy to gas species through electron-neutral momentum transfer collision process. According to our experimental conditions when the gas flow rate is lower than 2 $lpm$, heat is dominantly removed by conduction[29] through the capillary tube wall. While the gas flow rate is greater than 2 $lpm$, the effect of gas flow on cooling the $T_g$ and $T_{rot}$ becomes dominant through convection process which is proportional to the gas flow velocity. Hence $T_g$ and $T_{rot}$ decrease with increasing gas flow rate. On the other hand, $T_g$ and $T_{rot}$ increase with voltage, because electrons are collecting more energy from the increased electric field and transferring this energy to neutral particles through electron-neutral momentum transfer collision process. Thus, $T_g$ and $T_{rot}$ increase with increasing voltage but decrease with increasing gas flow rate. This result is found similar to that obtained by Ionascut-Nedelcescu *et al*. and Shou-Zhe Li *et al.*

It is also observed from Fig. 8 that $T_{rot}$ is slightly higher[8] than $T_g$. Taking into consideration the estimated errors for the determination of $T_{rot}$ and $T_g$, the two temperatures are different and disagrees the results obtained by Ionascut-Nedelcescu *et al*. This may arise due to the flow structure of the jet. Because, the flow dynamics with possible instabilities in the jet depend strongly on the conditions of the nozzle exit including initial velocity, Reynolds number and the nozzle geometry[7]. On the other hand, $T_g$ is determined using the $H_\beta$ line profile, while $T_{rot}$ is estimated from the Boltzmann plot using $OH$ band[30,31] may be the another reason for the difference. In our experiment however, the differences between $T_g$ and $T_{rot}$ are found within the considerable, both the experimental and analytical, limits.

### 3. Electron density

Interaction of ions and electrons are responsible[25] for Stark broadening, but electrons play significant role due to their lower masses and higher relative velocities. During OES data analyses, it is obtained that the contributions of $\Delta\lambda_{nat}$ and $\Delta\lambda_{res}$ are found negligible, but the dependence of $\Delta\lambda_{VdW}$ is on average $20\%$ on $\Delta\lambda_{L}$. Thus, the value of $\Delta\lambda_{VdW}$ is subtracted from $\Delta\lambda_{L}$ in order to obtain $\Delta\lambda_{Stark}$. However, the simplified form for the determination of $n_e$ using $H_\beta$ line profile is expressed[25] as

$$\Delta\lambda_{Strak} = 2.5 \times 10^{-10} \alpha {n_e}^{2/3} \ (nm), \qquad (3)$$

where $n_e$ is in $cm^{-3}$ and α is the ion broadened parameter. The values of α are obtained from the least square curve fitting of the Voigt profiles and found consistent with the tabulated data of Griem[32].

Figure 9 shows the effects of voltage and $O_2$ gas flow rate on electron density, where the symbols and lines represent the measured and calculated values of $n_e$, respectively. With increasing voltage, the electrons gain more energy from the enhanced electric fields and the ionizing collisions between electrons and neutrals become more frequent. Alternatively, the higher the value of electron temperature, the more effectively ionizes the neutrals causing abundant production of electrons. Therefore, most of the energy of the electrons, as gain from the electric field, being transferred to the oxygen species within the electrode gap through ohmic heating[27] and thereby producing more electrons and consequently $n_e$ increases with increasing voltage at constant $O_2$ flow rate.

On the other hand, the electron density increases as the $O_2$ gas flow rate increases. This dependence of electron density on gas flow rate and voltage is different from the findings of other atmospheric pressure plasma experiments[33,34] and similar to that of Moon *et al*. This can be attributed to different discharge conditions in that the speed of the gas fluid element is significantly different due to gas flow rate and diameter of the nozzle exit. The electric field is accumulated within the electrode gap of the plasma jet nozzle. Electron density at a distance $z$ in terms of gas flow rate in a simplified form, assuming that the velocity of the gas fluid element is uniform along the plume length, is expressed[27] as

$$n_e = \frac{16T_g}{\alpha_r z S T_{rot}}\xi\ , \qquad (4)$$

where $\alpha_r$, $z$, $S$, $\xi$ and $T_r$ are the three-body radiative recombination rate coefficient ($cm^3.s^{-1}$), axial position ($cm$) in the plasma plume length, cross-sectional area ($cm^2$) of the plasma plume, gas flow rate ($lpm$) and room temperature ($K$), respectively. It is seen from the above equation that $n_e$ is linearly proportional to the gas flow rate $\xi$. In the calculation, the value $\alpha_r = 2.85 \times 10^{-11} cm^3.s^{-1}$ is taken from Chung *et al*[35]. Within the experimental limit, the measured values of $n_e$ agree the calculated ones, as obtained using Eq. (4). Moon *et al.* obtained the similar results.

## 4. Excitation temperature

Excitation temperature is considered as the electron temperature in atmospheric pressure plasma[28]. For the determination of $T_{exc}$, Boltzmann plot is a simple and widely used method. In this method, the atomic transitions having large energy difference between the upper level ($E_j$) and the ground level are considered. Assuming that the atomic level population follows Boltzmann distribution, the emission intensity $I_{ji}$ can be expressed[36] as

$$I_{ji} = \frac{hcn_j A_{ji} g_j}{4\pi Z \lambda_{ji}} exp\left[-\frac{E_j}{k_B T_{exc}}\right], \qquad (5)$$

where, $\lambda_{ji}$, $A_{ji}$, $g_j$, $k_B$, $h, c, n_j$, $Z$ are the wavelength of the emitted light, transition probability of the level considered, statistical weight of upper level, Boltzmann constant, Planck's constant, speed of light, total population density and partition function, respectively. The Boltzmann plot for $\ln(I_{ji}\lambda_{ji}/A_{ji}g_j)$ is drawn as a function of $E_j$ introducing the parameter values required along with the measured intensity $I_{ji}$. $T_{exc}$ can be determined from the slope of linear fit. The upper and lower limits of $T_{exc}$ can be estimated in the similar way as mentioned earlier.

In this experiment, the spectral lines of $O^+$ at their emission wavelengths are selected. The transition parameters required for the determination of $T_{exc}$ are collected from the NIST database[21]. The Boltzmann plot and a linear fit are performed as shown in Fig.10, for an example. The effect of gas flow rate on $T_{exc}$ for different voltages is displayed in Fig. 11. It is observed that $T_{exc}$ decreases with increasing gas flow rate. This can be attributed with the Reynolds number as mentioned earlier. The Reynolds number is increased with increasing gas flow rate[7]. As can be seen from the relation of Reynolds number that the density of the oxygen molecules is increased with increasing the Reynolds number. In one hand, with increasing gas density, the mean free path is reducing that is electron-neutral collision frequency is increasing and hence electron excitation temperature is decreasing with increasing gas flow rate. On the other hand, with increasing applied voltage, electric field intensity in the electrode gap increases and hence electrons collect more energy from the enhanced electric field. As a result $T_{exc}$ increases with the increase of voltage. $T_{exc}$ in the plasma is almost linearly proportional to the electric field intensity. Uhm *et a.l*[37] contributed the similar results in their experiments.

## IV. Conclusion

Atmospheric pressure dielectric barrier discharge plasma jet is developed using capillary tube with oxygen as the working gas in order to generate nonthermal plasma for biological applications. Optical emission spectroscopy is used to diagnose the produced plasma. The spectral line broadening due to Doppler broadening and Stark broadening of $H_\beta$ line is used to determine $T_g$ and $n_e$. Consistent agreement between the calculated and measured values of $n_e$ is found. $T_{exc}$ and $T_{rot}$ are determined from $O^+$ line and $OH\ A - X\ (0,0)$ band of emissions employing Boltzmann plot method. It is found that $T_{rot}$ and $T_g$ decrease with increasing $O_2$ flow rate due to shorter transit time to transfer energy through electron-neutral collisions. $T_{exc}$ is also decreasing with flow rate due to increasing Reynolds numbers. So that $T_{exc}$ is decreasing with increasing gas flow rate. The effect of voltage shows that all the parameters are increasing with the increase of discharge voltage. Electrical measurement reveals that the discharge frequency decreases with increasing gas flow rate, but increases with applied voltage. Finally reactive oxygen species (ROS) of APDJ is applied on *Fusarium oxysporum* affected potato samples and it provides an initial effective result. Further study is going on to complete the expected application with this jet.

## ACKNOWLEDGEMENT

This work is supported by the Ministry of Science, Information and Communication Technology, Government of Bangladesh, under grant no.: 39.009.002.01.00.020.2010/EAS-07/766 and Faculty of Engineering, University of Rajshahi, under grants no.: 591/5/52/UGC/FE-9/2013 and 699(5)/5/52/RU/FE-4/2013.

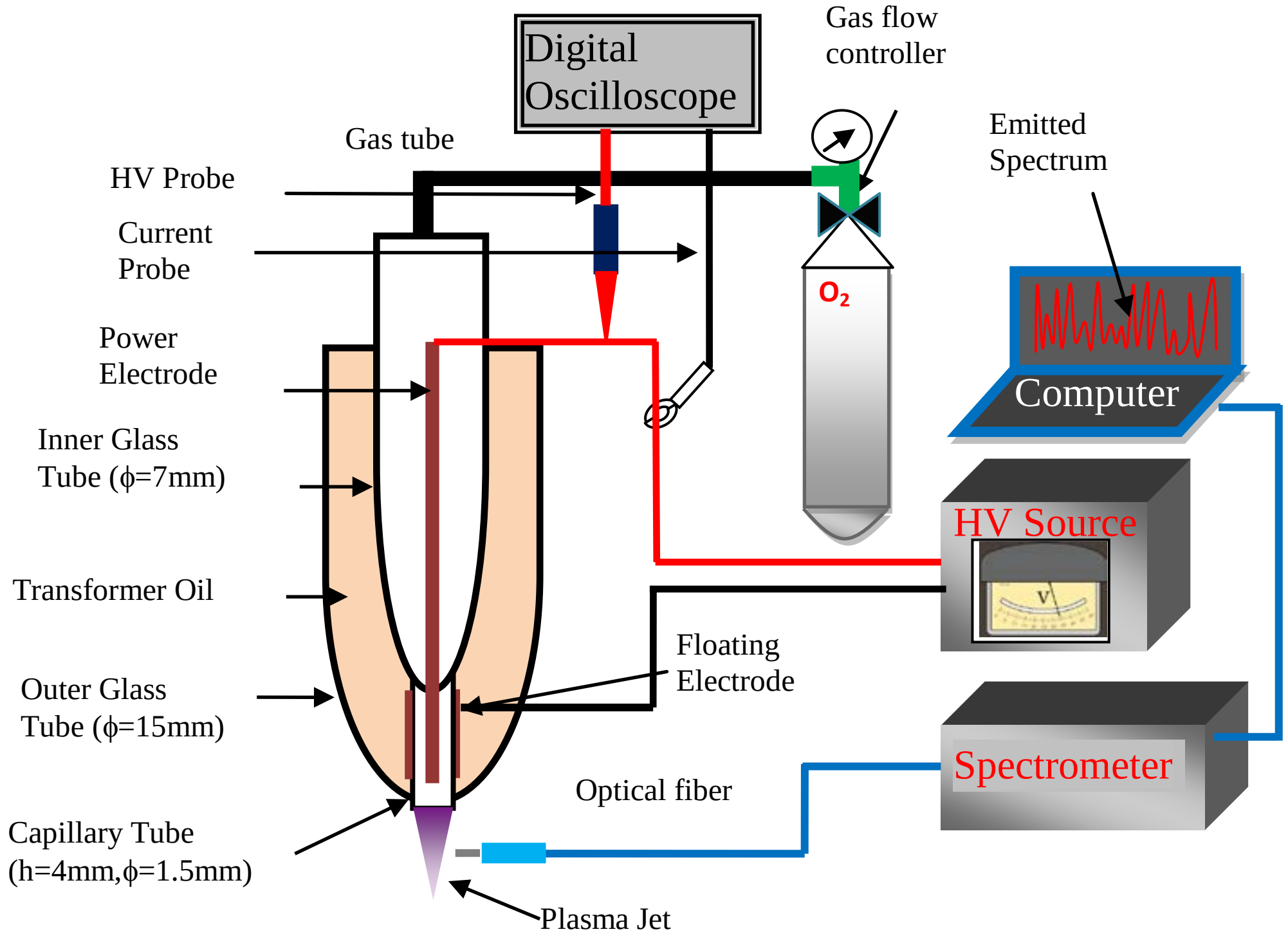

FIG.1: Schematic of the experimental setup of the atmospheric pressure capillary dielectric barrier discharge oxygen plasma jet.

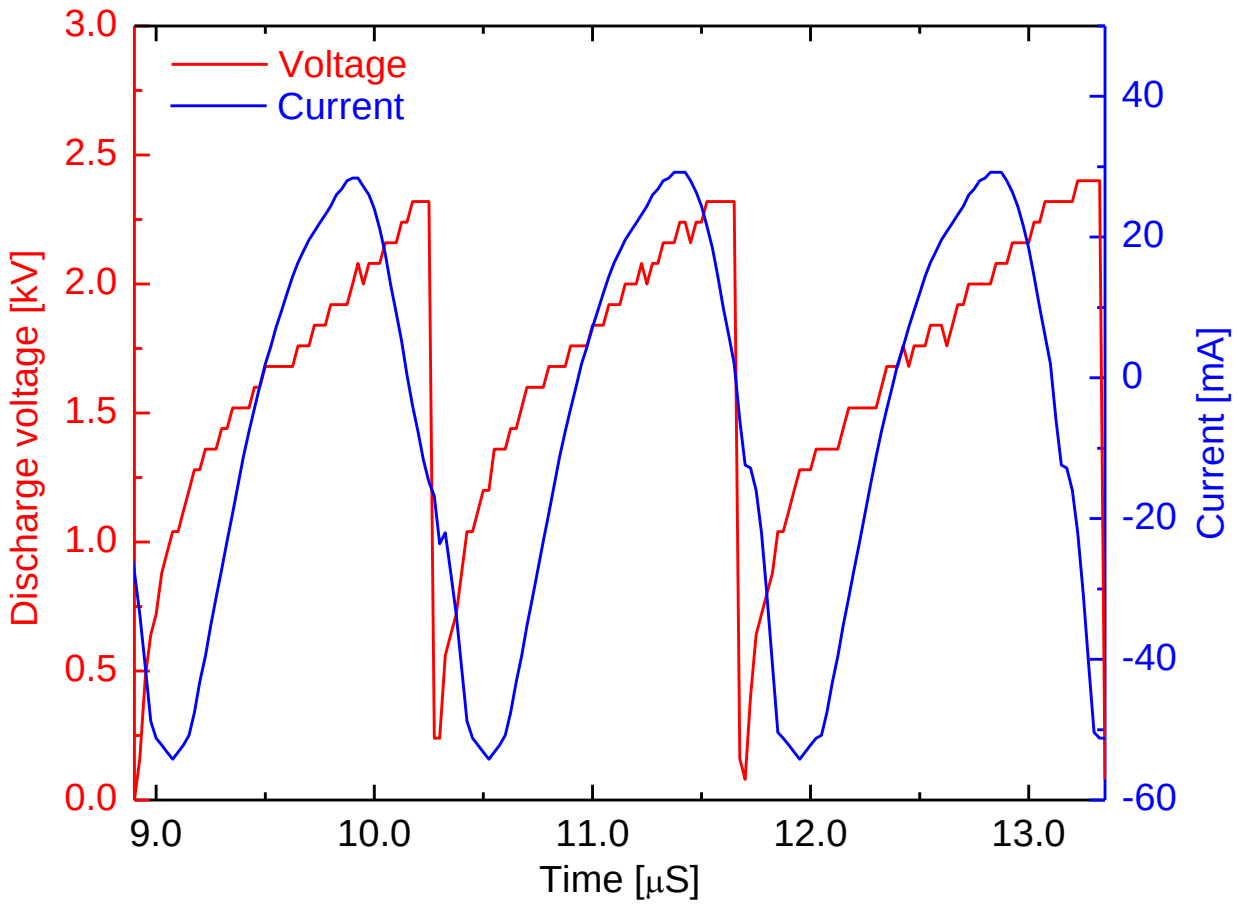

FIG. 2. Discharge voltage and current waveforms of the APDJ plasma measured at $2.5kV$ with $O_2$ flow rate of 3 $lmp$.

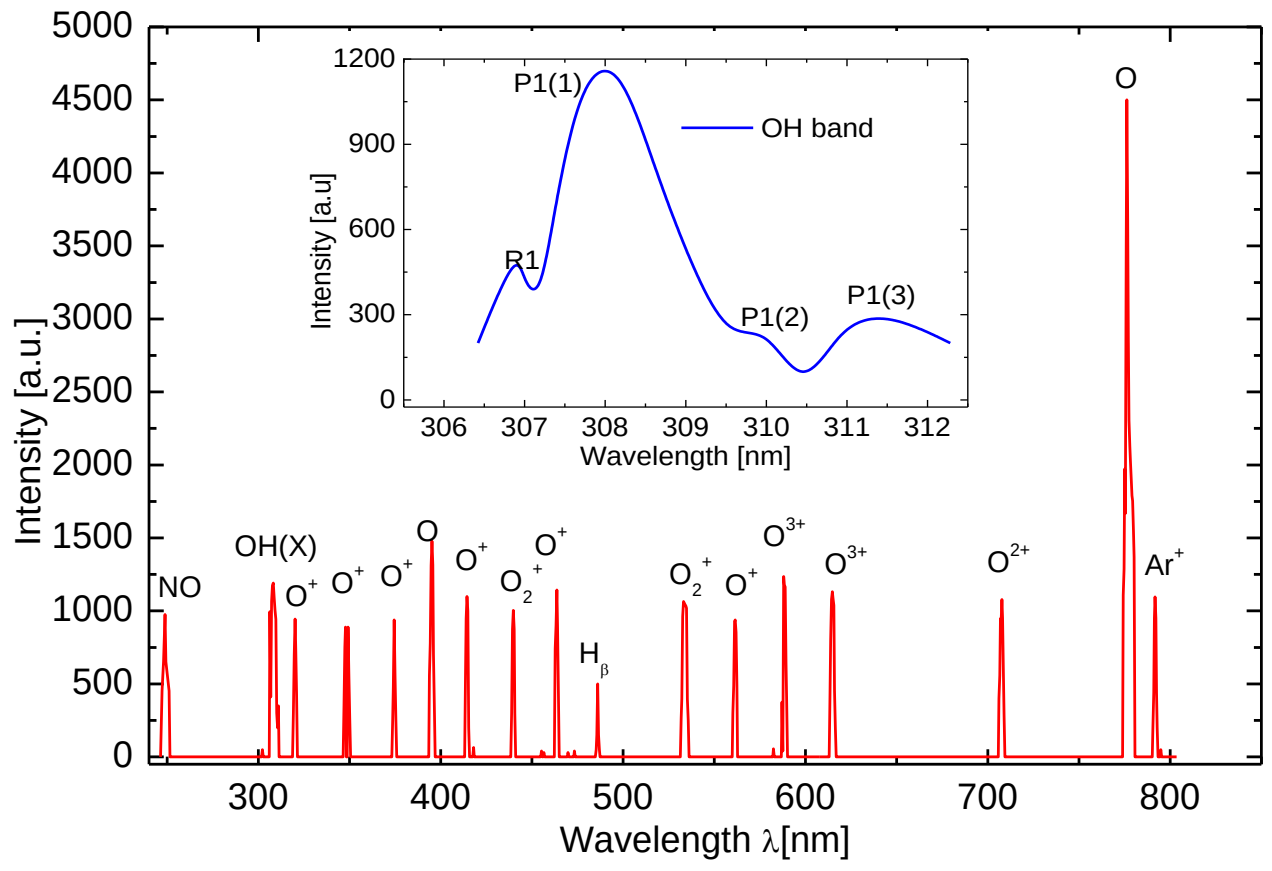

FIG. 3. Emitted spectrum recorded at $2.5 kV$ with $O_2$ flow rate of $3\ lpm$ along with the identified species. $R$ and $P$ branches of $OH$-band is shown in the inset.

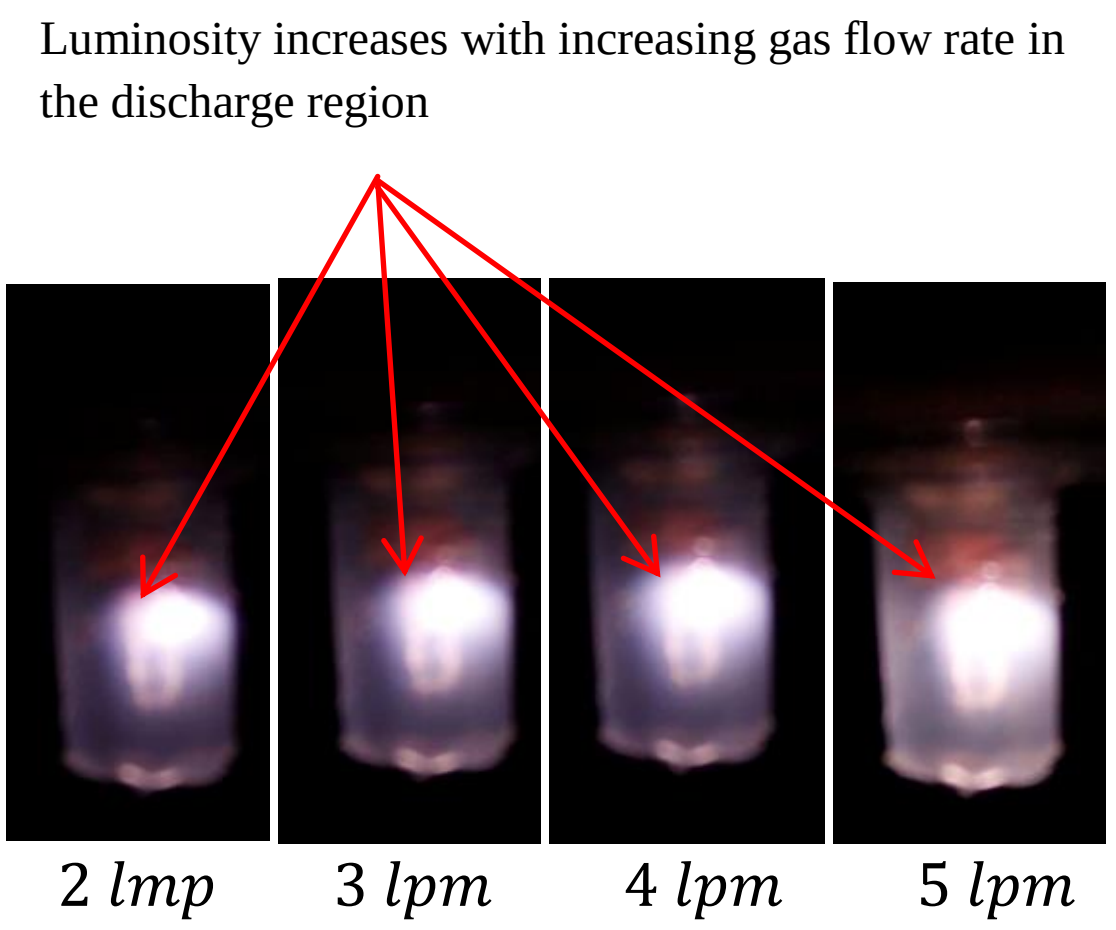

FIG. 4. APDJ discharge images are taken at 2.5 $kV$ as a function of $O_2$ gas flow rate.

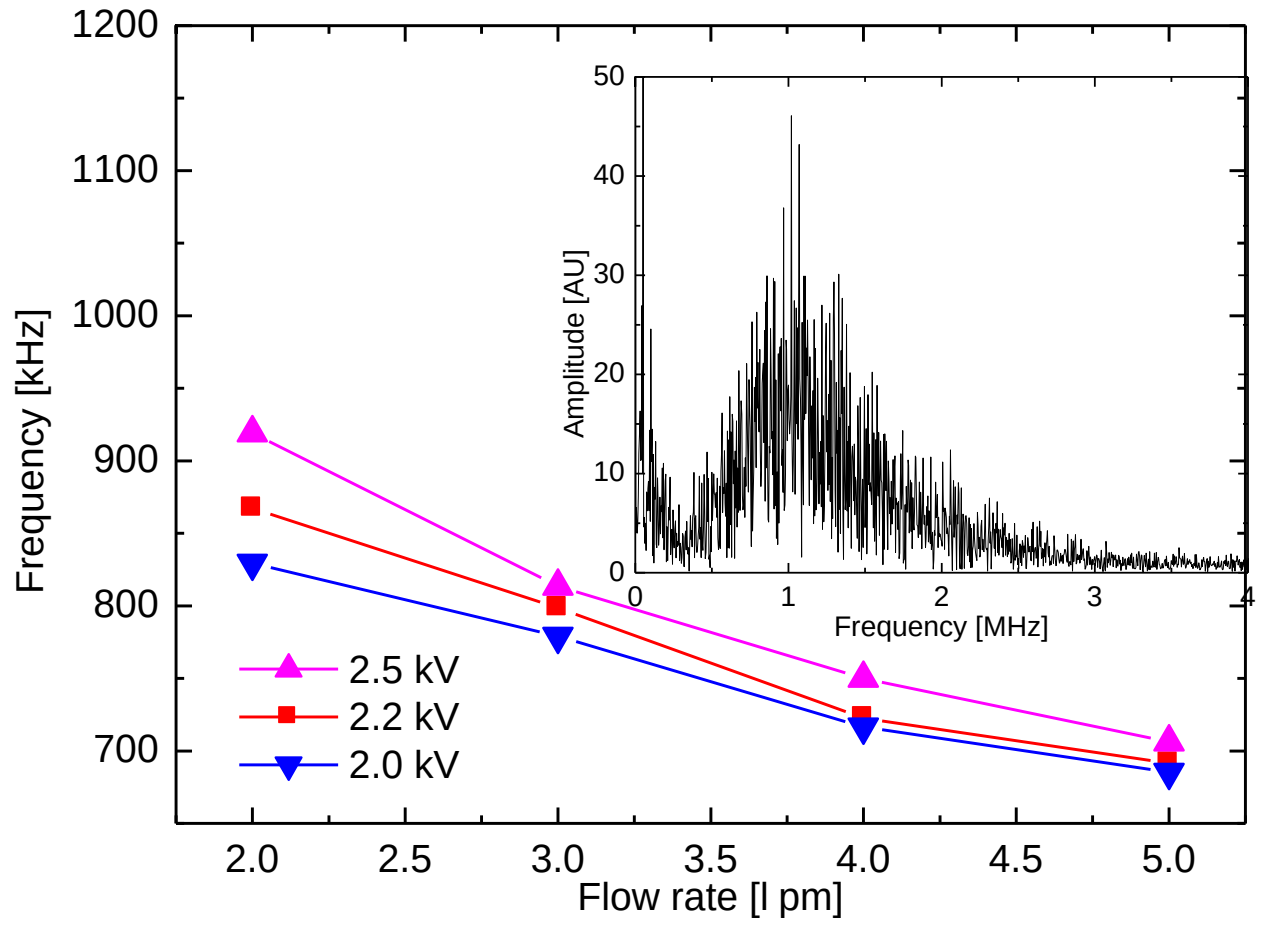

FIG. 5. Dependence of discharge frequency on $O_2$ gas flow rate as a function of voltage. Frequency spectrum is obtained by taking Fourier transformation of the discharge current measured at 2.5 $kV$ with $O_2$ flow rate of 2.0 $lpm$ as shown in the inset.

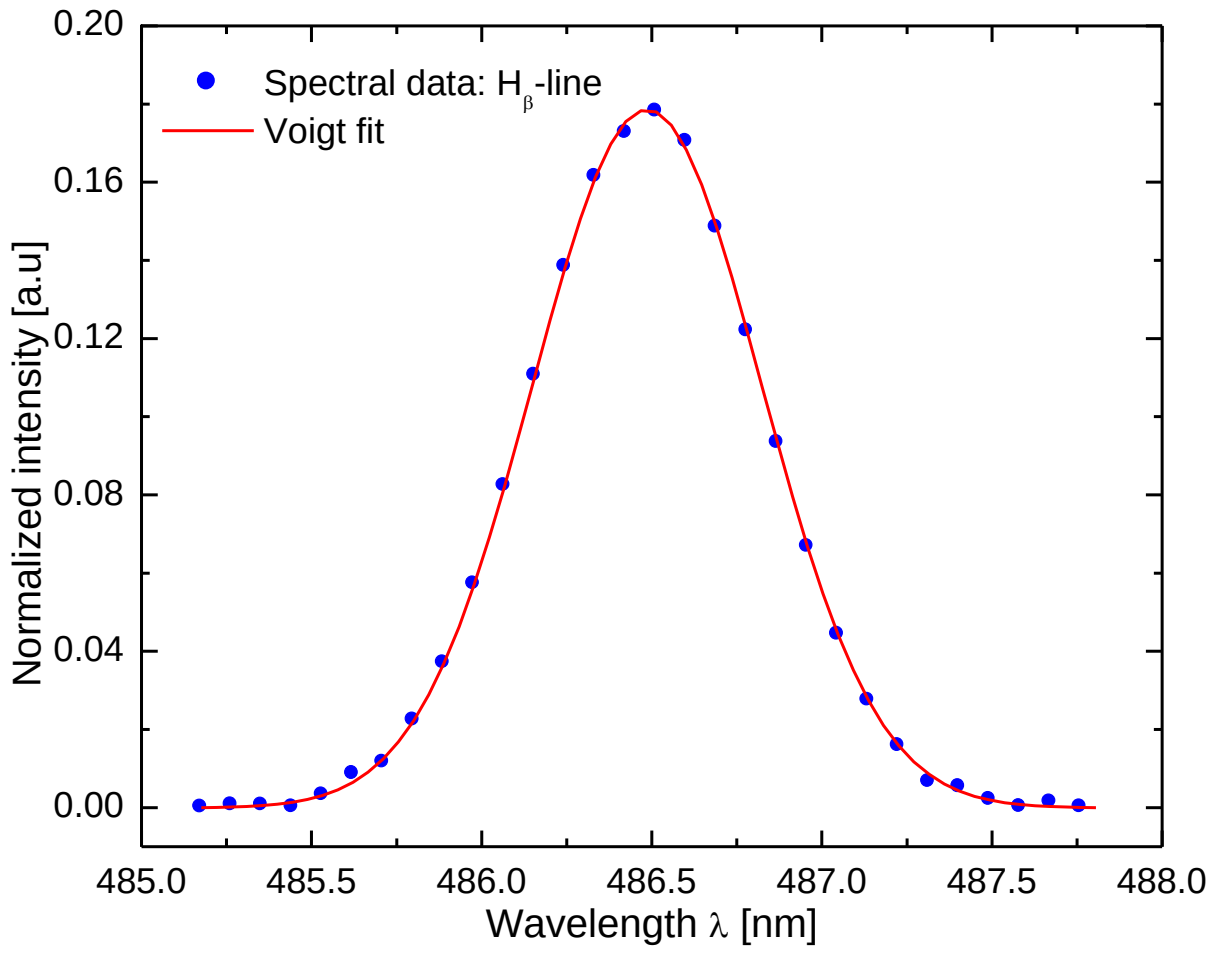

FIG. 6. Voigt function fitting of the spectroscopic data for $H_\beta$-line ($486.12\,nm$) measured at $2.5kV$ with $O_2$ flow rate of $3\ lpm$ to obtain broadenings.

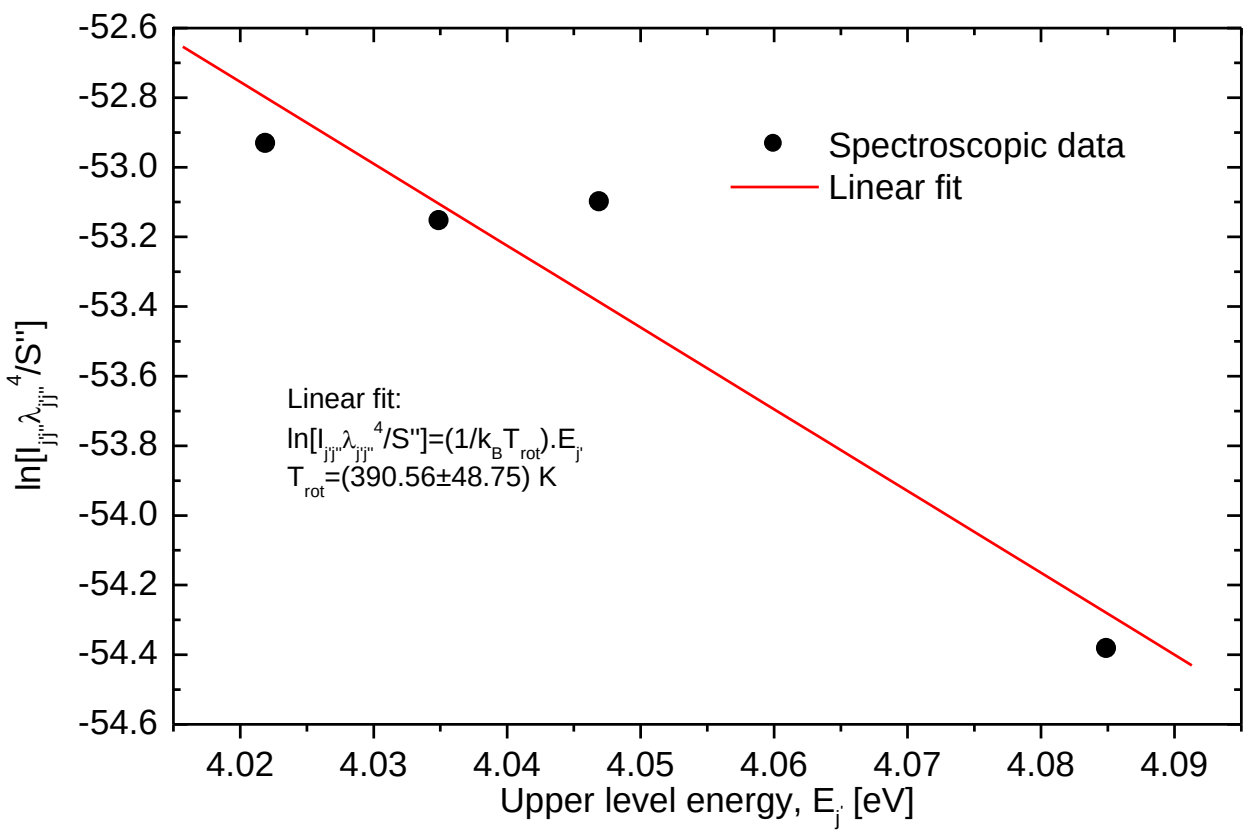

FIG. 7. Determination of rotational temperature $T_{rot}$ by the Boltzmann plot method using $306 - 312nm$ band of $OH$ radical (measured at $2.5kV$ with $O_2$ flow rate of $3\ lpm$).

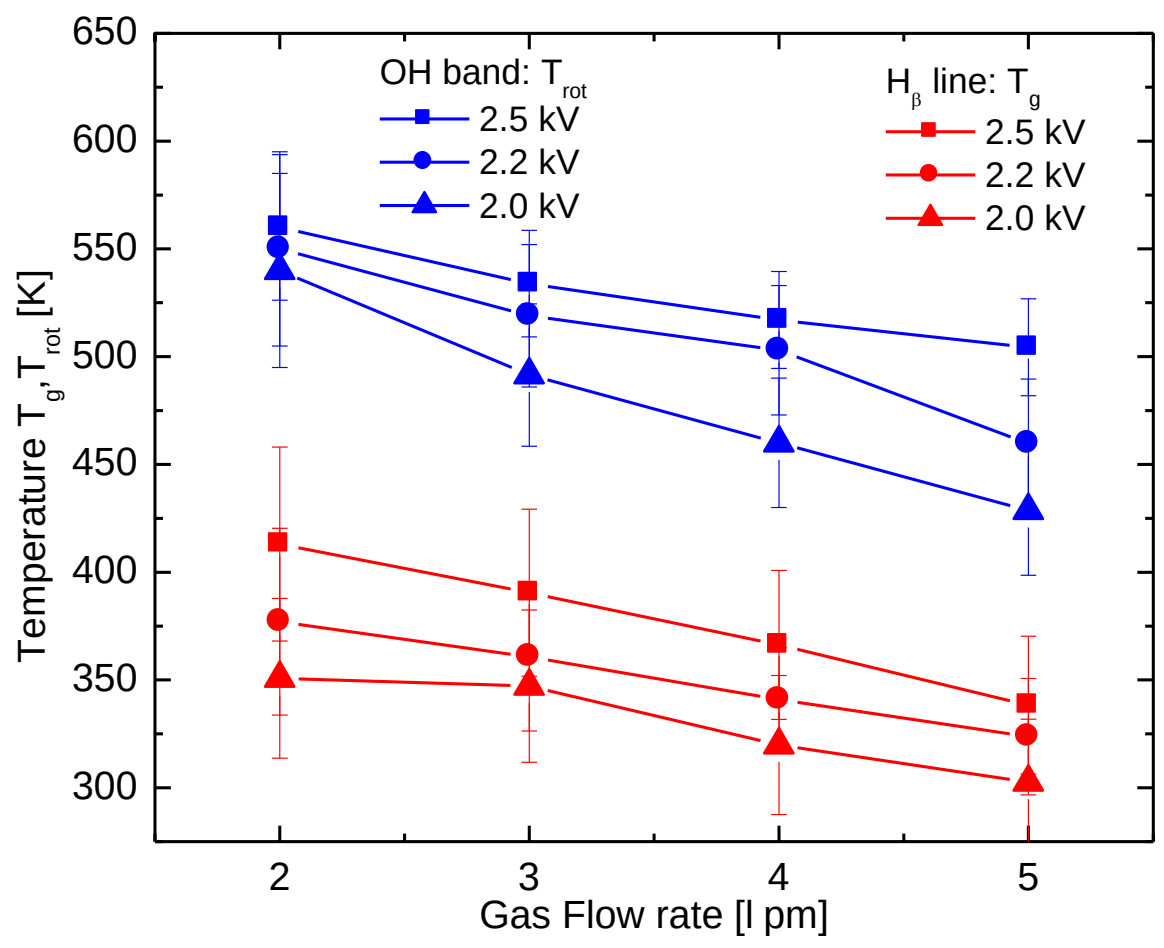

FIG.8: Dependence of gas temperature $T_g$ and rotational temperature $T_{rot}$ on $O_2$ gas flow rate and voltage.

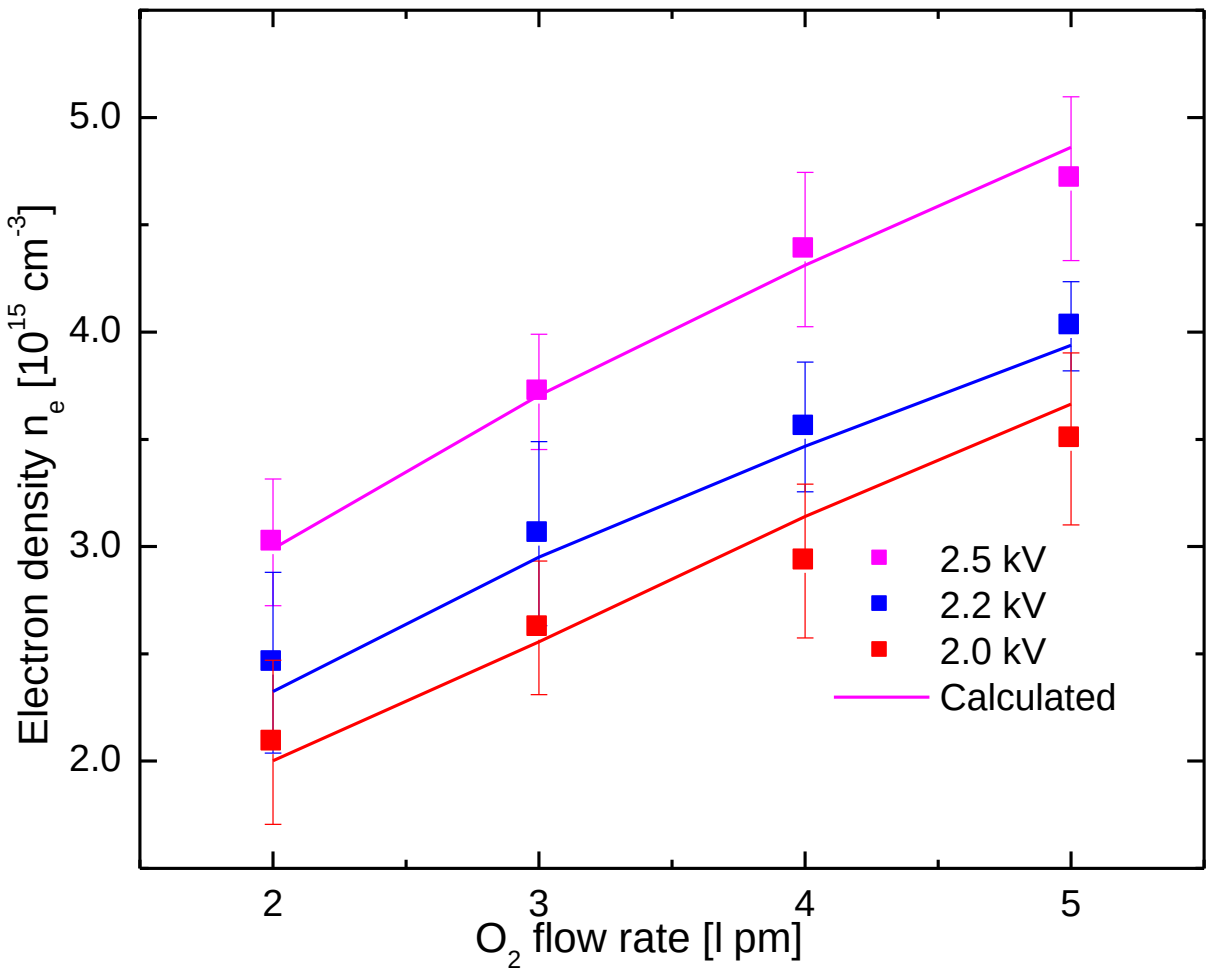

FIG. 9. Dependence of electron density $n_e$ on $O_2$ gas flow rate and voltage.

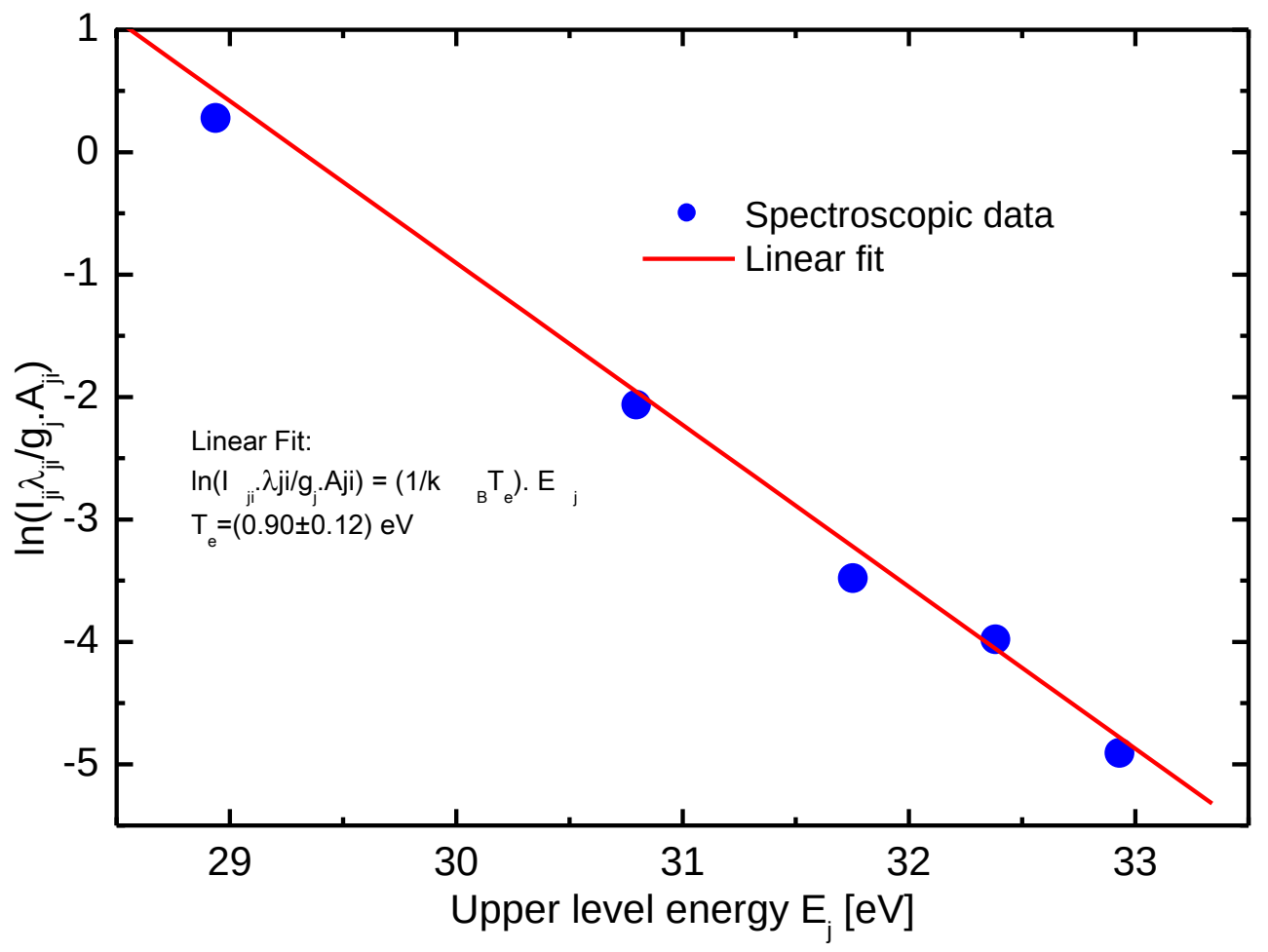

FIG. 10. Determination of excitation temperature, $T_{exc}$ by Boltzmann plot method using $O^+$ lines measured at $2.5kV$ with $O_2$ flow rate of 3 $lpm$.

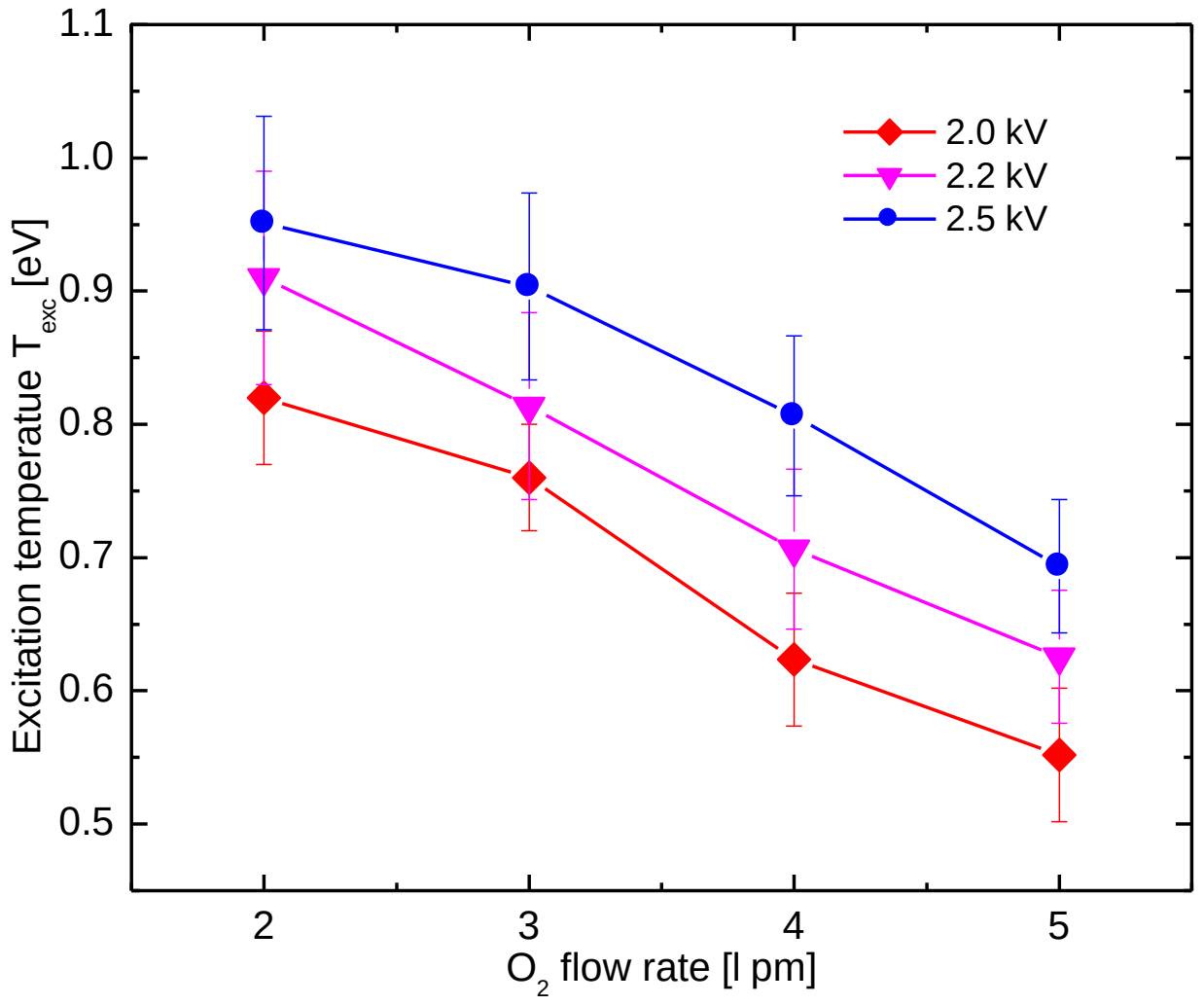

FIG.11. Dependence of excitation temperature $T_{exc}$ on gas flow rate and voltage.